**Knowledge Audit Framework (for Systems Engineering)**
**KAF-se**

**V1**

**10/12/2010**

Release: V1

Date: 2011

Catalogue entry

| Title | **Knowledge Audit Framework** |
|---|---|
| Creator | Institute of Socio Technical Complex Systems, Edinburgh UK |
| Subject | Knowledge Audit, Systems Engineering |
| Description | KAF consists of a process and some templates to guide the planning and execution of audits of knowledge resources, with emphasis on sharing.<br><br>KAF is based on methodological blueprint provided by the Data Audit Framework (DAF) conceived by the JISC-funded DAFD project.<br><br>KAF enables organisations to find out what knowledge resources are associated with the project, and how they are shared .<br><br>KAF is available in two versions<br>KAF-g (generic, domain independent )<br>KAF-se (targets systems enegineering knowledge) |
| Website | https://sites.google.com/site/kaframework/ |
| | |
| | |
| | |
| Language | English |
| Rights | |

**TABLE OF CONTENTS**

## Part 1. BACKGROUND TO THE KNOWLEDGE AUDIT FRAMEWORK

### a) Background Context and Overview

The Knowledge Audit Framework (KAF) is a technical instrument to support the systematic evaluation of Knowledge Sharing Artefacts and Practices adoption.

It is devised as part of a three years investigation into **knowledge sharing and reuse in systems engineering.**

**Currently KAF exists in two versions**
**KAF-g (a generic, domain independent version) and KAF-se targeted at the systems engineering domain.**
**This paper discusses te latter, KAF -SE v1**

KAF is designed to assist the evaluation of knowledge sharing practices in systems engineering, and it represents a portion (STEP 2) of the overall research project, illustrated in the diagram below.

A more extensive discussion of the research context is presented elsewhere. [1]

A field investigation into suitable methodological approaches to support and guide the development of a technical to address knowledge sharing and reuse challenges, [2]led to the discovery DAF (Data Audit Framework) recently published by HWAII at the University of Glasgow.

DAF is a published methodology and toolkit to audit data assets, and KAF has been modelled using DAF as a blueprint.

The rationale and motivation behind KAF pretty much fit the original statement: ‘a framework must be conceived to carry out an audit of departmental knowledge collections, awareness, policies and practice for knowledge curation and preservation.’ [3]

Similarly to DAF, KAF primary goal is to enable an objective evaluation of what knowledge resources are shared, where they are located and who is (or is not) responsible for them. KAF explicitly addresses the multidimensionality of Knowledge Reuse as a 'wicked problem'[4] and places particular emphasis on the knowledge sharing and reuse.

---

1 https://sites.google.com/site/kaframework/
2 DAF website
3 reference
4 wicked problem

Knowledge Audits consist typically of the following components: an Inventory, mapping and a knowledge flow.[5]

The current version of KAF includes a template for knowledge audits developed specifically with the knowledge requirements for the systems engineering domain, as explained in the relevant section [6] It is envisaged that additional templates can be developed to suit other domains in future work.

Finally, a set of methodological notes justify the framework development choices where KAF differs from DAF[7]

## b) KAF Vs DAF

KAF is a knowleddge auditing framework, modelled on DAF. It is shorter and simpler than the original methodology which is to be considered as its source reference, however where possible it preserves some of its core characteristics, namely:

- general methodological approach
- structure (the KAF process is articulated in four stages)
- some of the original arguments are retained, and where suitable, portions of the original methodology are adopted
- the Dublin Core element set representing the essential metadata for each asset are retained, a mapping of KAF template elements to Dublin Core is provided *see Annex 3)

**KAF differs from DAF as follows:**

- KAF targets knowledge resources
- SCOPE : DAF is aimed primarily at Higher Education institutions, while the current version of KAF is intended to be used primarily at to audit 'projects': an institution can undertake many projects, and a knowledge resources are audited for each project. Future versions may follow a different direction.
- PROCESSES: like DAF, KAF is articulated around four stages, however each step in the corresponding KAF stage are different from DAF: for example, KAF does not require on site visits and

5 knowledge audits reference
6 Paragraph/section
7 See Paragraph methodological annotation full access to project documentation, but relies on knowledge that is shared primarily on the web and by electronic means of communication, and that can be accessed (or not accessed) via inspection of the project website, or remotely via electronic communication exchanges with the project team (mainly emails).

- AUDIT MODE: KAF is designed to be remote audits, its templates can also be used during onsite visits if desired
- KAF rationale is based on a different conceptual/taxonomic set of relations than the original DAF
- KAF emphasises the audit of Knowledge Sharing artefacts and procedures in relation to their reusability

To ensure maximim fidelity to the original DAF methodology, portions of the structure and wording of this documentation and paper replicate the original where possible

## c) Introduction

KAF provides guidance and technical instruments to plan and execute an audit of knowledge resources taking into account the multidimensionality of the knowledge sharing problem space, and to help to form an objective assessment of the degree of sharedness of knowledge resources. It rests on the following central tenets:

- Knowledge generated by publicly funded research should be shared in as much as possible, consistently with research funding bodies policy, where available, and other constraints (a more detailed policy evaluation is being drafted in related work) KAF targets 'explicit knowledge', intended as :

  *As third and intermediary type, explicit knowledge is seen as an interface for human interaction and for the purpose of knowledge externalisation, the latter one ending up in external knowledge.* [8]

- The knowledge asset inventory templates adopted in this version of the framework can be amended and modified to target specific knowledge models of the intended target domain in future work

## d) Aim and Scope of the Knowledge Audit Framework

[8] Model based Taxonomy for Knowledge Development Scenarios http://www.iaeng.org/publication/WCE2010/WCE2010_pp289-294.pdf

Publicly funded research generates 'knowledge', however no mechanism exist to assess  exactly what knowledge, and what effort is necessary to access and reuse this knowledge.The KAF methodology contains a template and a process to carry out 'knowledge resources audits', that can be applied in principle to any project, organisation or institution, although the current working version targets the systems engineering domain, developed with EPRSC grant for Networked Capabilities in Systems Engineering (Nectise). The original DAF methodology relied upon a concept map that uses two taxonomic groups for data assets (‘by origin’ and ‘by nature’), as indicated in the original DAF document.[9]  KAF by contrast takes is shaped taking into account the organisational knowledge taxonomy (declarative and procedural, causal)[10]

Source: Jose Vasconcelos, (modified)

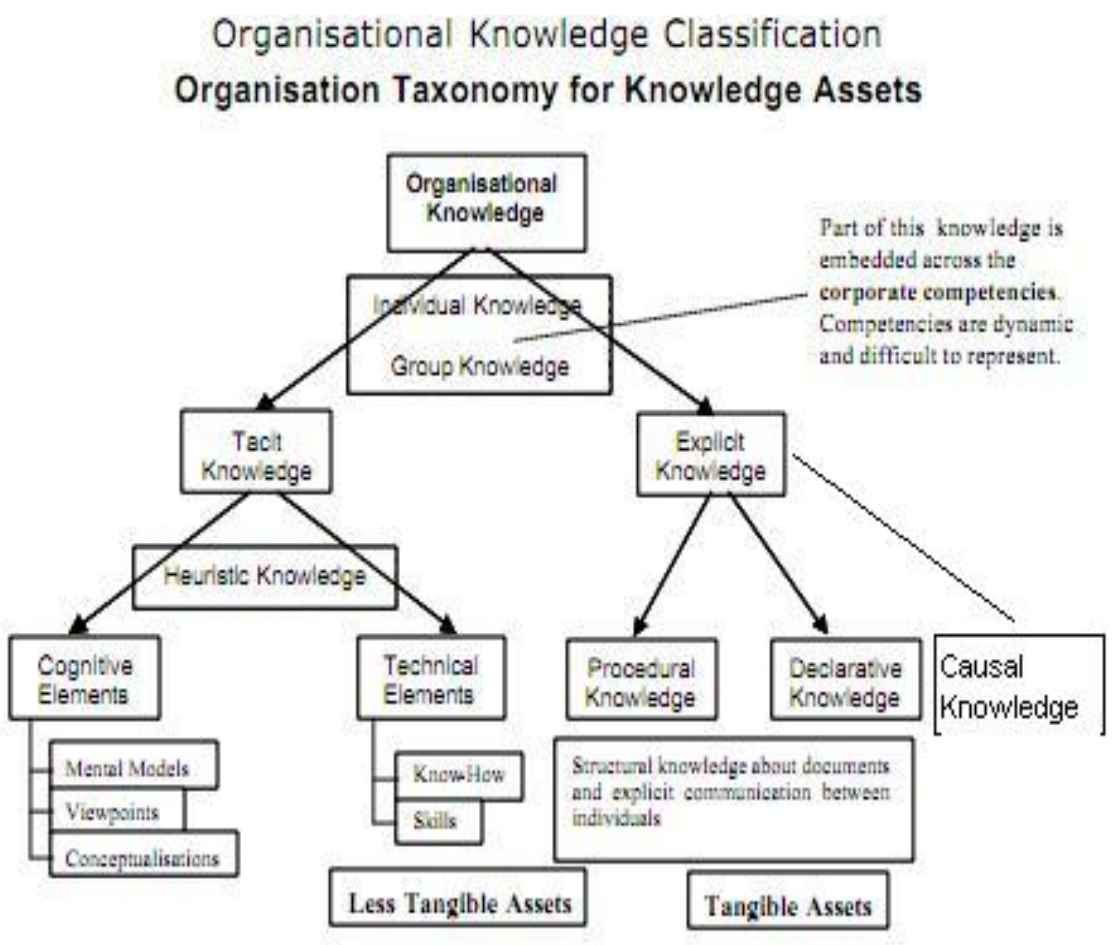

### e) Some Methodological Considerations for the KAF Template

The current version of the KAF methodology presented in this document, has been developed as part of a research into Networked Enabled Capabilities in Systems Engineering (NECTISE, EPRSC ); as such, it is designed to target knowledge, in particular the auditing of knowledge sharing artefacts in the systems engineering domain.

Among other elements, KAF differs from DAF, as highlighted in the introductory notes of this document, by using a different template more suited to auditing 'knowledge resources', the preferred term in our research. Knowledge resources are well known in literature as 'knowledge assets [11]

---

[9] DAF doc, reference paragraph

[10] Organisational Knowledge Taxonomy, http://www2.ufp.pt/~jvasco/RWT-02-presentation.pdf
Vasconcelos, J.B., Kimble, C. and Gouveia, F.(2000), A Design for a Group Memory System using Ontologies. Proceedings of 5th UKAIS Conference, Cardiff, Wales, McGraw Hill, ISBN: 0077095588, pp. 246 - 255.

[11] http://www.eurojournals.com/irjfe_21_04.pdf

Resource in this research is preferred to asset. as the latter has commercial implications empahsising the economic value that can be obtained through the exploitation of knowledge as intellectual property By contrast, our research emphasises the availability of knowledge free from commercial implications.

Nonaka and Takeuchi group Knowledge Assets into four categories, two tacit and two explicit, as illustrated in the image below. ( Image)

The two explicit assets categories directly relevant to KAF, reflected in the design of the knowledge inventory template design[12] are:

A - Systemic Knowledge Assets, Combination, Systematizing (Virtual collective): Explicit, codified, systematic, descriptive, complete, comparative, evaluative. EXAMPLE: DOCUMENTS, SPECIFICATION, MANUALS

B – Conceptual Knowledge Assets, Externalization, Originating (Face-to-face individual): Symbols, concepts, brands, styles, metaphors, analogies, emergent, developmental. EXAMPLE: PRODUCT CONCEPTS, DESIGNS

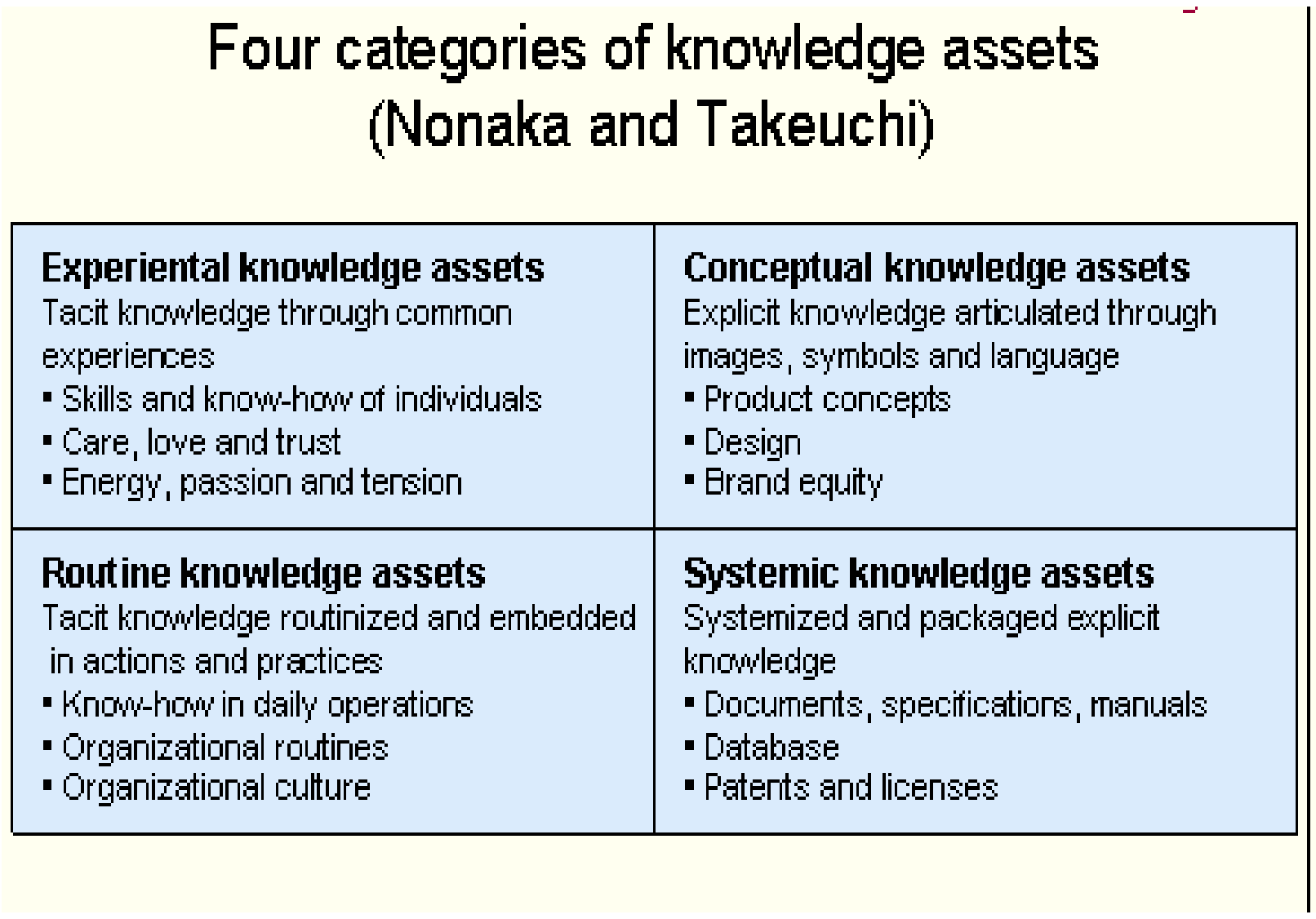

Four categories of knowledge assets (Nonaka and Takeuchi)

| | |
|---|---|
| **Experiental knowledge assets**<br>Tacit knowledge through common experiences<br>▪ Skills and know-how of individuals<br>▪ Care, love and trust<br>▪ Energy, passion and tension | **Conceptual knowledge assets**<br>Explicit knowledge articulated through images, symbols and language<br>▪ Product concepts<br>▪ Design<br>▪ Brand equity |
| **Routine knowledge assets**<br>Tacit knowledge routinized and embedded in actions and practices<br>▪ Know-how in daily operations<br>▪ Organizational routines<br>▪ Organizational culture | **Systemic knowledge assets**<br>Systemized and packaged explicit knowledge<br>▪ Documents, specifications, manuals<br>▪ Database<br>▪ Patents and licenses |

The next paragraph provides a summary of main arguments and considerations that justify the modelling choices of the KAF template structure (APPENDIX NR)

12 De Geytere, T. (2005) "SECI model (Nonaka Takeuchi)" [Web Page]. URL http://www.12manage.com/methods_nonaka_seci.html [2005, October 29].

## f) **Engineering Knowledge Structures for Systems Engineering**

The American Engineers' Council for Professional Development (ECPD, the predecessor of ABET)[1] defined "engineering" as:

> [T]he creative application of scientific principles to design or develop structures, machines, apparatus, or manufacturing processes, or works utilizing them singly or in combination; or to construct or operate the same with full cognizance of their design; or to forecast their behavior under specific operating conditions; all as respects an intended function, economics of operation and safety to life and property.[2][3][4]

### The term 'knowledge engineering'

**Knowledge engineering** (KE) was defined in 1983 by Edward Feigenbaum, and Pamela McCorduck as follows:

> KE is an engineering discipline that involves integrating knowledge into computer systems in order to solve complex problems normally requiring a high level of human expertise.[1

In the absence of a general ontology for Systems Engineering (the development of which is being discussed in relevant communities at the time of writing) practitioners [13]can summarise knowledge pertaining to the systems engineering domain into broad 'upper categories' , for example:

states
entities (objects)
processes (transformations)
axioms (rules)

A useful term of reference for what constitutes 'knowledge structure' in a technical domain is provided by Romizowski's Knowledge Schema [14] a simplified version of which is reproduced as a diagram below

---

[13] INCOSE SSWG Discussion, Dov et al
[14] Romizowski

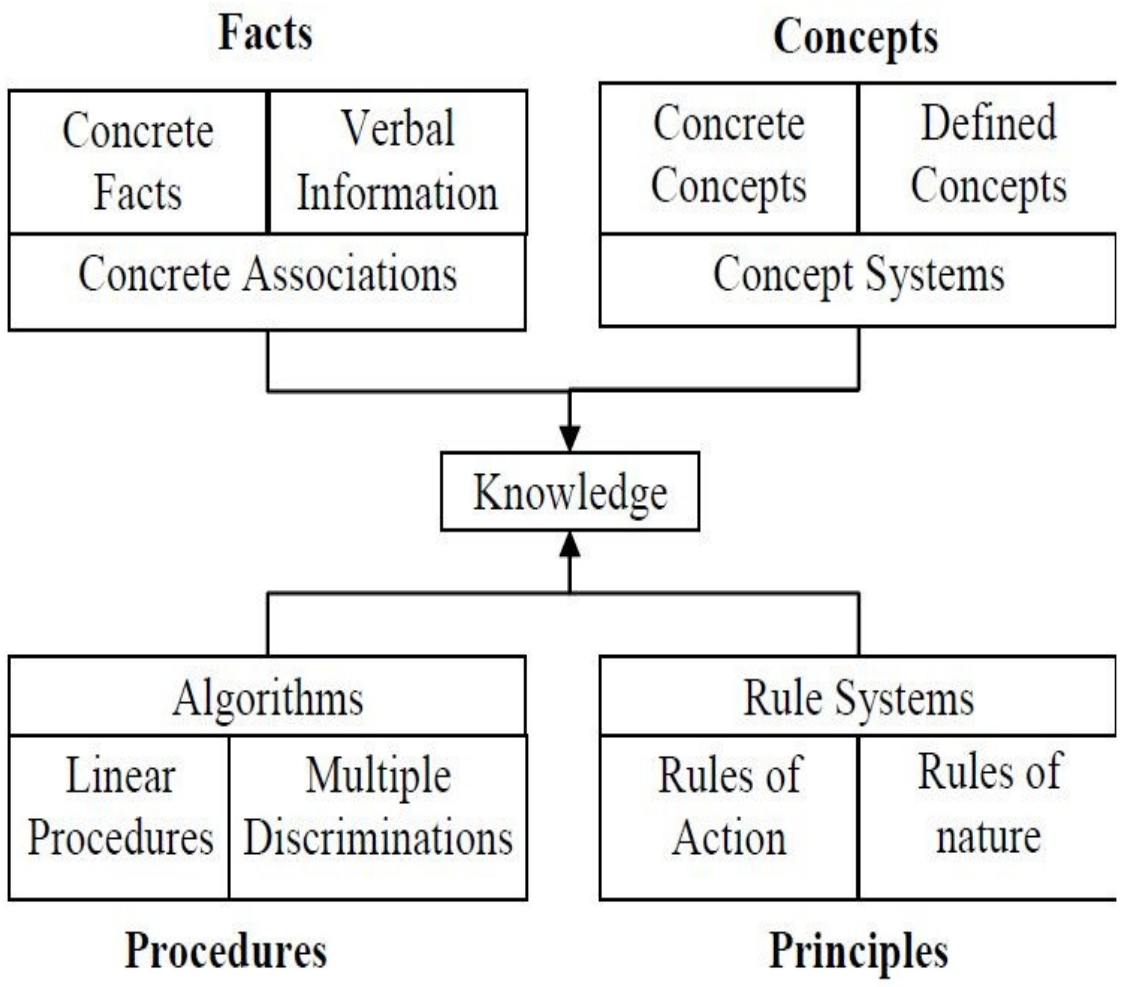

Modified Romiszowski Mechanics Framework (MRMF)

Romizowski categorization, described in more detail in the table below, in KAF is adopted loosely as a means of reinforcing the distinction between FACTS and CONCEPTS as '*declarative knowledge*', and ALGORITMS and RULES as '*procedural knowledge*'.

Additionally, where available KAF adds 'provenance' (from Dublin Core Element Set) as 'causal knowledge', thereby extending the Vasconcelos taxonomy of organisational knowledge (image nr)

| Category | Sub-category | Definition and example |
|---|---|---|
| Facts | Concrete Facts | Things committed to memory from simple observations, and not associated with language. Eg. remembering someone's face, recognition of an object |
| | Verbal Information | Knowledge associated with language or symbols. Eg. units, terminology, vector notation etc. |
| | Concrete Associations | Interlinking of facts. Eg. recognizing a truss analysis problem, knowing which given quantity is velocity etc. |
| Concepts | Concrete Concepts | Simple concrete facts tied to understanding. Eg. recognizing a cantilever beam |
| | Defined Concepts | More complex verbal and factual information tied to understanding. Eg. Knowing that a vector has magnitude and direction and the associated terminology |
| | Concept Systems | Interrelated concepts. Eg. momentum is a product of mass and velocity which in turn require understanding. |
| Procedures | Linear Procedures | Simple, chain calculations. Eg. substituting numbers into an equation and solving. |
| | Multiple Discriminations | Distinguishing between information, and solving problems in parallel. Eg. knowing/deciding which numbers to substitute into an equation. |
| | Algorithms | Complete procedures involving both linear procedures and multiple discriminations. Eg. Truss analysis where several problems need to be solved simultaneously using the correct data and processes. |
| Principles | Rules of Action | Rule's governing the behaviour or actions of the individual. Eg. identifying all given information a the start of a problem solution. |
| | Rules of Nature | Rules that explain the behaviour of objects or the surrounding environment. Eg. Gravity is what pulls objects down, forces cause the motion of objects. |
| | Rule Systems | Strategies and theories. Eg, a particular approach to solving a large problem. |

Declarative = glossary, lexemes corresponding to facts

Procedural = rules.

Causal = we add a causal dimension – indicating provenance and sourcing, for example, according to what policy is knowledge shared? What is the source of knowledge and references/citations?

**PART 2**

**2) Motivation: The Need for a Knowledge Audit Framework**

a) Why Audit Knowledge ? From Data to Knowledge
b) How the Knowledge Audit Framework can Help
c) Dimensions of KAF (cognitive, organisational, technical)
a) Why Knowledge Assets?

a) Why Audit Knowledge?

KAF targets primarily explicit knowledge , which is knowledge which has been 'codified' in the form of cognitive artefacts, referred to as 'knowledge assets'. [15]
A view of the relation between data, information and knowledge is provided in the image below[16]

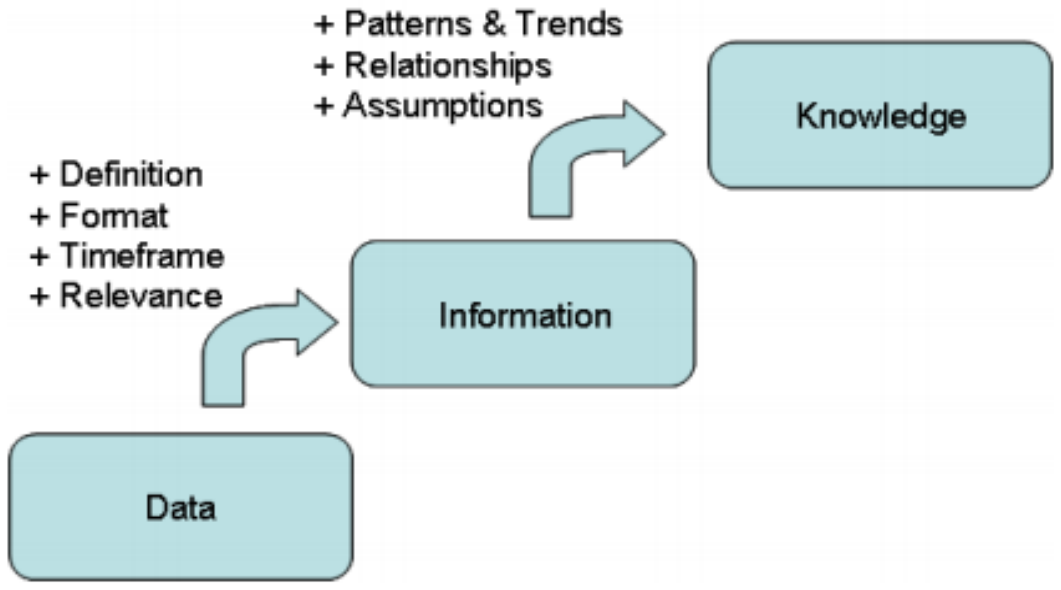

KAF extends DAF by targeting more articulate cognitive artefacts than elementray 'data' The continuum data-information-knowledge is well established in literature[17] and illustrated in the image above.

In developing a suitable Knowledge Audit for the Systems Engineering domain, key notions from the knowledge auditing practice are adopted [18]

15

16 http://media.techtarget.com/searchDataManagement/downloads/A_taste_of_DAMA_DMBOK.pdf
17 .Ackoff, R. L., "From Data to Wisdom", Journal of Applies Systems Analysis, Volume 16, 1989 p 3-9.

18 Debenham and Clark (1994)

According to literature[19] :

*A knowledge audit is a planning document, which provides a structural overview of a designated section of an organization's knowledge as well as details of the qualitative and quantitative characteristics of the individual chunks of knowledge within that designated section. The document also identifies the knowledge repositories in which those chunks reside. They feel that the knowledge audit is a scientific measurement of the state of affairs of specified sections of corporate knowledge. A critical part of aknowledge management methodology is performing a knowledge audit* (Liebowitz,1999).

A knowledge audit can include any of the following components:

- Knowledge Needs assessment
- Knowledge Inventory
- Knowledge Flow
- Knowledge Map

**Knowledge analysis methods that could be used in the knowledge audit[3]**

| Sl.No. | Knowledge analysis methods | Usage in Knowledge Audit |
|---|---|---|
| 1. | Questionnaire-based knowledge surveys: | To obtain broad overviews of an operation's knowledge status |
| 2. | Middle management target group sessions | To identify knowledge-related conditions that warrant management attention |
| 3. | Task environment analysis | To understand, often in great detail, which knowledge is present and its role |
| 4. | Verbal protocol analysis | To identify knowledge elements, fragments, and atom |
| 5. | Basic knowledge analysis | To identify aggregated or more detailed knowledge |
| 6. | Knowledge mapping | To develop concept maps as hierar chies or nets |
| 7. | Critical knowledge function analysis | To locate knowledge-sensitive areas |
| 8. | Knowledge use and requirements analysis: | To identify how knowledge is used for business purposes and determine how situations can be improved |
| 9. | Knowledge scripting and profiling | To identify details of knowledge intensive work and which role knowledge plays to deliver quality products |
| 10. | Knowledge flow analysis | To gain overview of knowledge exchanges, losses, or inputs of the task business processes or the whole enterprise |

[3] *Based on Wing (1993)*

The current version of KAF includes a template and a process to perform **Knowledge Inventory** with emphasis on knowledge sharing artefacts, and taking into account the structures of technical knowledge. Future versions of KAF will include Knowledge mapping and knowledge flows.

A Knowledge Audit is a technical instrument to appraise knowledge sharing practices: given different challenges, and taking into account the nature of the problem KAF is designed to address, the current deliverables consist mainly of an inventory template that targets systems engineering knowledge with particular emphasis on knowledge reuse.

---

19 Liebowitz

## b) How Does the Framework help?

To effectively manage knowledge reuse, an organisation must adopt knowledge sharing artefacts and behaviours. Conducting an audit enables a systematic mapping of knowledge reuse artefacts, highlighting blind spots and weaknesses in the Knowledge Sharing and Reuse behaviour, and show the gaps to improve overall Knowledge Management strategy for the research projects, and for the organisations that initate and deliver them.

The framework is designed to be used both at project and organisational level without dedicated or specialist staff and with limited investment of time or effort.

The current version of the audit templates is designed to collect information required to evaluate the level of adoption of knowledge reuse artefacts and practices. The audit addresses five core questions:

1. What knowledge resources for each publicly funded systems engineering project in the UK are shared, therefore publicly reuseable?
2. Where are these assets located?
3. What knowledge sharing mechanism/techniques are adopted (or not)?
4. Who is responsible in the organisation for making knowledge shareable?
5. Is the organisation adopting/following a knowledge sharing and reuse policy?

The information collected by KAF provides a clear overall picture of organisational knowledge sharing practice. Organisations armed with this information can make changes to improve existing knowledge management accordingly. KAF provides a simple method of collecting and using this information. The following chapter outlines how to use KAF and details additional sources of support.

## c) Knowledge sharing and reuse dimensions

In related research [20], the entangled complexities of knowledge sharing and reuse are considered a 'wicked problem'. Some of the dimensions of this complexities are identified as follows:

**Cognitive** Language, level of skill required to use/adopt the knowleddge

**Organisational** Policies and management practices that promote knowledge

[20] WIMMS 2010 Paper

sharing and reuse

**Technical** Choice of codification, knowledge representation techniqes, and standards adopted

The knowledge audit template (see annex) is designed to capture the diversity of dimensions and factors that make up knowledge reuse in systems engineering in the UK as discussed in the relevant sections of this methodology paper.

**PART II THE KNOWLEDGE AUDIT FRAMEWORK**

KAF consists of:

a) An Audit process
b) An Audit template
c) An online repository to collect and analyse the data of the audited project
d) Examples of forms for communicating with project teams and their organisations
e) Guidance for using the knowledge Audit Framework

## a) The Audit Process

Audits are carried out remotely, and via email and other means of remote communication including telephone. Each audit is expected to take between 1-3 days (8- 25 work hours) from start to finish.

There are four main stages to the knowledge Audit Framework, illustrated in the image below, and explained in the following section:

Note: this diagram needs updating

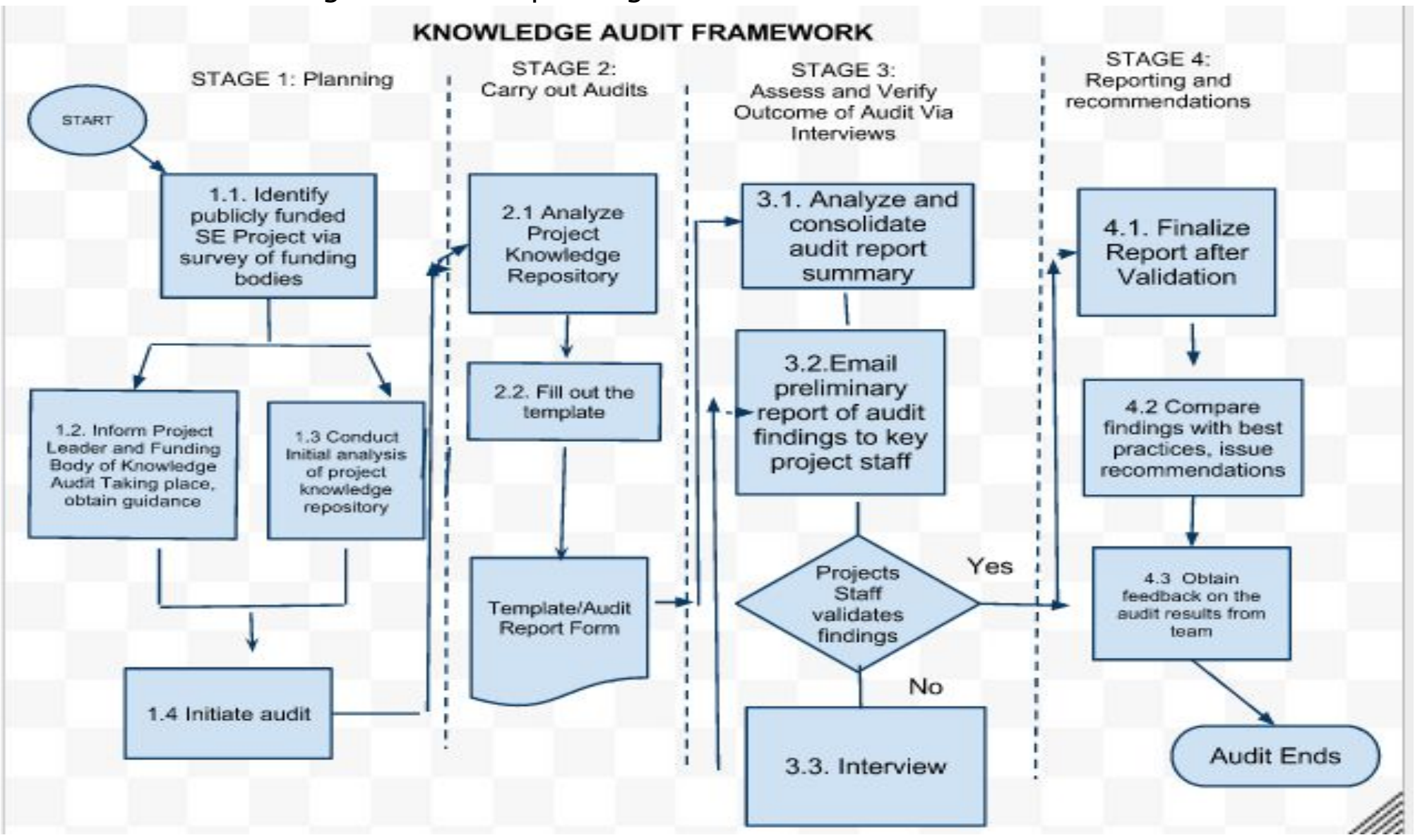

IMAGE: The four stages of the Knowledge Audit Framework

**Stage 1: Planning (identify project and key team members)**

1.1 Identify the project (survey all funding bodies in the UK for publicly funded systems engineering projects and select a meaningful subset)
1.2 Inform project leade about KAF procedure being undertaken, ask for input if required/useful
1.3 Conduct initial analysis of the repository
1.4 Initiate Audit

**STAGE 2 : Execution (carry out the audit)**

2.1 Analyse Knowledge Resources for the Project
2.2 Fill out the slots in the audit form template, possibly using the online data entry tool

**STAGE 3: Verification**
3.1 Consolidate results of the audit in a report
3.2 **Send a copy of the initial report to the project team for validation**
3.3 **if** not valid, **then** carry out an interview **and** amend report
(repeat 3.3 untiil 3.2 is validated OK)

**STAGE 4: Reporting findings and recommending change.**
4.1 After inventory summary is approved by the project lead, finalise the report
4.2 Compare report findings with good practices, issue recommendations
4.3 Get feedback from the team on the KAF process

Audit Ends
--------------

## B) The template

KAF includes a template for the inventorying of knowledge resources.

A generic knowledge inventory template can fit most knowledge auditing requirements, since knowledge resources are well defined in literature, as discussed above, but no generally reusable guidelines exist for their detailed specification.

KAF Knowledge Asset Inventory template however is designed to contribute to fulfilling the need for a specification for knowledge resources Reuse in Systems Engineering, and serve as a guideline for specifying knowledge inventory template in other domains.

In this section some of the justfication are presented that underpin the design choices of the KAF inventory template, bearing in mind that future and altenative versions can be modified to suit different purposes.

A set of knowledge resources can be scoped according to different criteria, for example as identified in paragraph 'KAF Scope' in this paper, a distinction can be made between 'systemic' and 'conceptual', which can be translated into procedural and declarative knowledge respectively.

KAF also incorporates in part the notion of project assets dimensions proposed by Chourabi et al [21] who describes the main facet of a "SE-Project Asset" as the central concept for SE project knowledge modeling.

-**Domain facet:** contains basic concepts and relations for describing the content of engineering assets on a high semantic level. (domain ontology)
-**Product facet:** contains concepts and relations representing artifact types as well as their information model. In SEdomain, a system is described with several views such as: contextual, dynamic, static, functional or organic. By formally
relating modeling elements to domain concepts we could provide a systematic and semantic description of an,engineering solution.
-**Process facet**: contains concepts and relations that formally describe engineering activities, tasks, actors, and design rationales concepts (intentions, alternatives, argumentations and justification for engineering decisions).

The image below shows how to integrate Nonaka's and Chourabi views of knowledge resources, leveraging their complementarity:

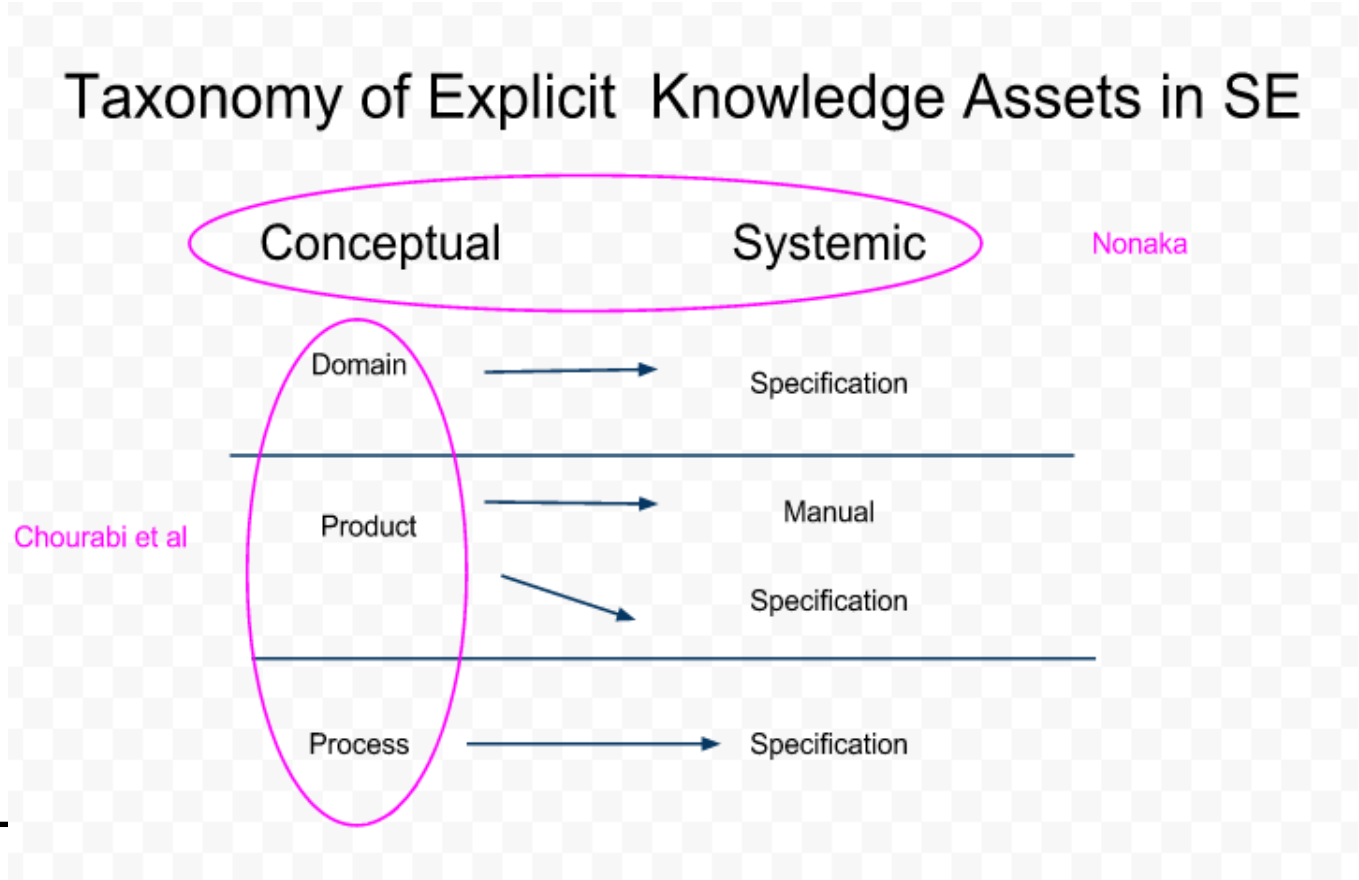

KAF integrates the dimensions supplied by Nonaka and Chourabi by making explicit their complementarity: the conceptual and systemic knowledge resources are complementary, and that different degrees of formalization (doumentation, manuals, specification) are desirable for each different category of conceptual asset, for example Domain, Product and Process.

[21] Ontology Based Knowledge Modeling for System Engineering Projects Olfa Chourabi, Yann Pollet, Mohamed Ben Ahme

KAF implicitly postulates that '*to be shareable, to each conceptual asset should correspond a systemic one'.*

Typically, the most detailed knowledge representation of a system is the 'System Specification' , which can be high level and generic, or very detailed with a high level of granularity. No overall uptodate standard exists for 'system specification' to the best of authors knowledge.

The main KAF audit template therefore aims to reflect at least in part standard system specification outline, which is made up of different areas and components[22] An summative example of system specification is provided below:

**A SYSTEM SPECIFICATION CAN INCLUDE:**

**1.0 System Overview** (natural language)
Goals and objectives
System statement of scope
Main requirements
A description of the entire system. Major inputs, processing functionality and outputs are described without regard to implementation detail.
System context (The system is placed in a business or product line context. Strategic issues relevant to context are discussed. The intent is for the reader to understand the "big picture.")
Major constraints (Any business or product line constraints that will impact the manner in which the system is to be specified, designed, implemented or tested are noted here)

**2.0 Functional and Data Description** (natural language, structured language, diagram)
This section describes overall system function and the information domain in which it operates.
System architecture (diagram)
A context-level model of the system
Subsystem overview (diagram)
Data Description /Model
Top-level data objects that will be managed/manipulated by the system or product are described in this section.
Major data objects (diagram)
Relations (diagram, ERD)
External Interface Description (natural language, diagram)
The system's interface(s) to the outside world are described.
Internal interfaces
Human interface (UML)

**3.0 Subsystem Description** (for each subsystem) (text)
Description for Subsystem n (diagram)
Subsystem scope (TEXT DOCUMENT)
Subsystem flow diagram
Subsystem n components *for each component (diagram, text)

22 www.rspa.com/docs/**Systemspec**.html

Another possible approach to developing a knowledge artefact inventory template to facilitate reuse, is to map knowledge artefacts to the system development lifecycle: system development lifecycle is typically envisaged as 'phases', undertaken iteratively as needed (until the requirements are met) throughout the life of the development [23]

Image SE Lifecycle (ref 20)

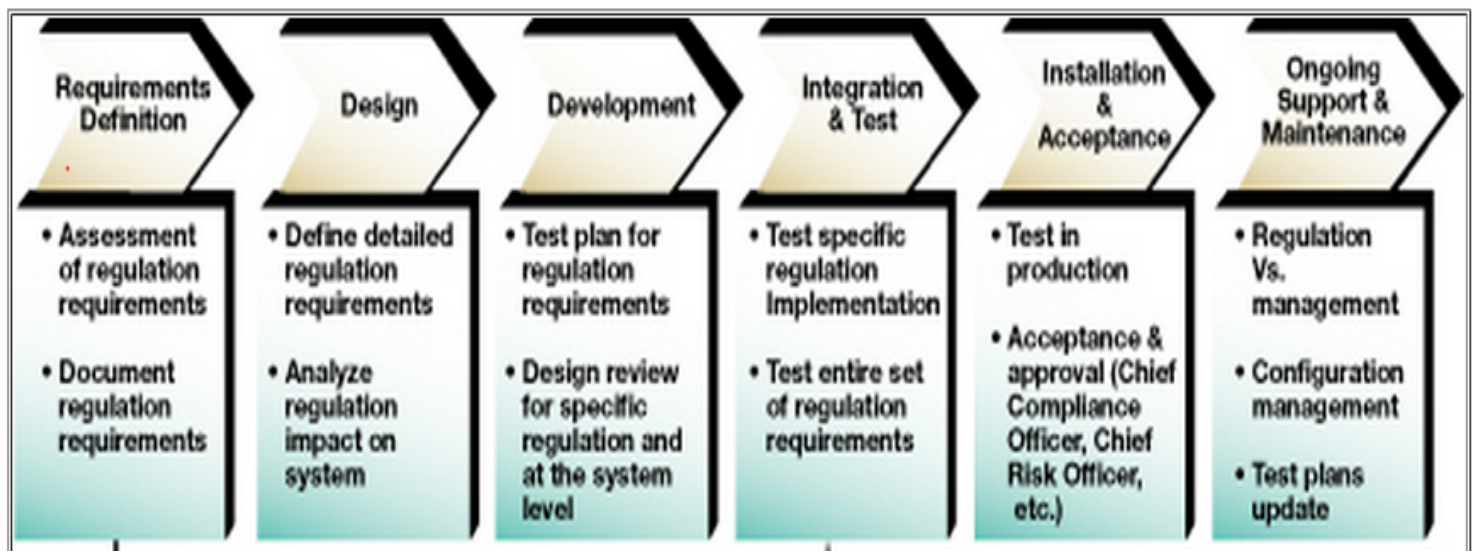

To each system development phase, a number of technical document correspond that can be qualified as 'knowledge resources'

| | | |
|---|---|---|
| requirements | >>>>>>>>>>> | specification document (narrative) |
| design | >>>>>>>>>>> | diagram, narrative |
| development | >>>>>>>>>>> | system specification (narrative, diagrams) |
| integration | >>>>>>>>>>> | interface specification, standards |
| testing | >>>>>>>>>>> | test plan |
| installation | >>>>>>>>>>> | operating manual, user guide |
| acceptance | >>>>>>>>>>> | contractual agreement |
| support | >>>>>>>>>>> | user feedback, tickets |
| maintenance | >>>>>>>>>>> | feedback, new requirements |

The KAF template is constructed to refect and integrate various taxonomic representation and views of knowledge resources in relation to a) standard system specification b) lifecycle phases (See Appendix 1)

The following table summarises the lifecycle phases and corresponding knowledge resources, and mapping format, notation formalisms and commonly adopted sharing mechanisms for each.

**TYPES OF knowledge resources IN SYSTEMS ENGINEERING**

| LIFECYCLE PHASE | KNOWLEDGE ASSET *document, specification | FORMAT | NOTATION/FORMALISM | SHARING MECHANISM |
|---|---|---|---|---|

23 The Impact of Regulation on Information System Planning J. Nadivi Internal Auditor, 2009 http://www.theiia.org/intAuditor/itaudit/2009-articles/the-impact-of-regulation-on-information-system-planning/

| | | | | |
|---|---|---|---|---|
| analysis | requirements specification | narrative<br>structured text | natural language,<br>pseudocode | image<br>word document<br>spreadsheet<br>pdf<br>html<br>xml<br>rdf<br>owl<br>other |
| design | system diagram | diagram | ER, DF, UML | |
| development | system specification | narrative<br>structured text | Natural language<br>pseudocode | |
| installation | operating manual<br>user guide | narrative<br>diagrams | Natural language<br>graphics | |
| testing | test plan | structured text | natural language<br>charts | |
| acceptance | contract | narrative | natural language | |
| support | user feedback<br>tickets | narrative | natural language | |
| maintenance | feedback<br>new requirements | narrative | natural language | |

**LIFECYCLE INDEPENDENT knowledge resources**

| |
|---|
| **Process** |
| **Ruleset** |
| **standard compliance document** |

## c) The online data collection tool

A prototype online collection form for the collection of data for each project, is provided as part of this deliverable for the purpose of this study.

**http://tinyurl.com/5rg4wyj**

It is implemented using 'Google apps' spreadsheet and form.
It is articulated in three sections as follows:

- About the Project
- About each Knowledge Asset
- Summary KA Report for each project

## d) Examples of Communication

Annex 2 provides sample templates to communicate with project team as follows:
a) a note to inform the funding body of audits taking place, describing the purpose, scope and process

b) a note to inform the project team/ leader and obtain input and guidance
c) an email to the project team leader to verify/validate the correctness of the report containing preliminary findings
d) a note to communicate the final findings and issuing recommendations

**e) Guidance Notes: How to Use the Audit Framework**

This methodology is based around audits being conducted at project or organizational level. KAF audits can be performed indepedently and remotely, and do not depend on the cooperation or time availability of project team member. However the resulting audit report can be validated or amended by project team members in stage 3. discussed in more detail below.

The working version of the KAF template is structured as follows

1. Information about the project being audited
   (name of the project, funding body, key team members, duration, website, publications)
2. Inventory of shared knowledge, articulated to specify declarative and procedural knowledge, across cognitive, organisational and technical dimensions.

KAF process and template may be developed iteratively based on the feedback received during the audits.

PART III VALIDATION AND CONCLUSIONS, FUTURE WORK

## Validation

A more detailed discussion of the methodological and validation challenges of wicked socio technical problems is presented in related publications.

The KAF metodology is considered valid if,
After performing the knowledge asset audit, it becomes possible to answer one or more of the following initial questions:

1. What knowledge resources for each publicly funded systems engineering project in the UK are shared, therefore publicly reuseable?
2. Where are these assets located?
3. What knowledge sharing mechanism/techniques are adopted (or not)?
4. Who is responsible in the organisation for making knowledge shareable?
5. Is the organisation adopting/following a knowledge sharing and reuse policy?

The validation mechanism is therefore 'heuristic' and (empirical?)

Four (4) pilot cases and forty (40) audits are planned as validation mechanism for this version of KAF, the analysis of which is going to be published in papers

## Conclusions, Acknowledgements and Future Work

This paper introduces KAF, a knowledge audit methodology, which consists of an auditing process and a template, and associated online collection tools.

The template provided in this version of KAF is designed to audit knowledge artefacts generated as deliverables of publicly funded projects in the systems engineering domain.

Future versions of the framework will include different templates, designed to inventory knowledge resources in different domains, as well as various degrees of refinement.

This research is partly supported by EPRSC grant nr ….

Special thanks to HATII Team members for cooperation and permission to reuse parts of the DAF methodology.

KAF website
https://sites.google.com/site/kaframework/

Contact:
Paola Di Maio,
KAF Principal Investigator
paola.dimaio@gmail.com

c/o DMEM
University of Strathclyde
Room 106m 75 Montrose Street
Glasgow UK
paola.dimaio@strath.ac.uk

## APPENDICES

Appendix 1: Template audit forms and worked examples
a) Audit Form Part 1: Audited organization
b) Audit Form Part 2: Inventory of knowledge resources

Appendix 2: Guidance documents
a) Example email approach to target organization

Appendix 3: Mapping KAF element set to Dublin Core

Appendix 4: Glossary

# APPENDIX 1

## TEMPLATE

## a) About the organization

**PART 1. ORGANISATION**

| | | OTHER COMMENTS |
|---|---|---|
| PROJECT NAME | | |
| DESCRIPTION | | |
| URL/PROJECT DOCUMENT/WEBSITE | | |
| PARTNERS | | |
| FUNDING BODY | | |
| KS POLICY | | |
| CONTRACTUAL CLAUSES THAT IMPACT KS OF THIS PROJECT | | |
| | | |

## b) For each knowledge resource

**2. FOR EACH KNOWLEDGE RESOURCE OF THE PROJECT**

| | |
|---|---|
| RESOURCE NAME/ID | |
| RESOURCE TYPE | |
| DESCRIPTION | |
| MANTAINED BY | |
| LAST UPDATED | |
| NEXT REVIEW DUE | |
| LANGUAGE | |
| DOES THIS RESOURCE COMPLY WITH ANY STANDARD? | |
| IS THE USE OF THIS RESOURCE PRESCRIBED BY AY POLICY? | |
| FORMAT | |
| LICENSE | |
| URL | |
| OTHER LOCATION | |
| DOES THIS RESOURCE REQUIRE ANY PERMISSION TO BE ACCESSED/REUSED? | |
| | |
| LIFECYCLE PHASE | |

**APPENDIX 2**

a) a note to inform the funding body of audits taking place, describing the purpose, scope and process

Dear *person at funding body

I am writing to inform you that I am carrying out a study that involves performing project audits of publicly funded research.

The rationale and motivation and methodology are listed on this publicly accessible website:

https://sites.google.com/site/kaframework/

The projects funded by your funding body (insert name) which we intend to audit are enclosed in a list attached.

Each project leader will be contacted in the next few days to be informed of the audit procedure, which takes place remotely and unobtrusively via searches and via the respective project websites, so that they can point us to relevant repositories and sources of knowledge that can be publicly evaluated.

A copy of the summary findings will be emailed to you as soon as available.

Please do let me have any questions you may have at this stage

Best Regards

Your Signature

**b) Example letter  project leader**

Dear *project team leader

I am writing to inform you that I am carrying out a study that involves performing project audits of publicly funded research.

The rationale and motivation and methodology are listed on this publicly accessible website://sites.google.com/site/kaframework/

The projects being audited are enclosed in a list attached.

The audit procedure takes place remotely and unobtrusively via searches and through the project website, but in case the some of the information the audit aims to identify cannot be easily accessed on your website, I would be most grateful if you could provide it by filling out the relevant portion of the attached form (for example, are there project team members in charge of knowledge sharing, so that any further questions can be directed to them)

Also feel free to point me to relevant repositories which are not listed on the project website, and that we can include when performing the inventory.

I enclose the preliminary information gathered about your project in a form enclosed

A copy of the summary findings will be emailed to you as soon as available, so that you can approve them or correct them, and then a final version of the inventory will be included in a public audit report.

Do not hesitate to ask questions you may have at this stage

Best Regards

Your Signature

Encl 1. List of projects being audited
Encl 2. Preliminary Project information obtained from funding body

**c) An email to verify the preliminary finding**

Dear *Project Leader Name

Following our email dated …
I am writing to inform you about the findings of the knowledge inventory carried out on your project, as discussed.

The findings are enclosed in the following summary

- name of project, project details
- name of person in charge of KM
- number of publicly available knowledge resources for this Project

Please do let me know if the above is correct, or please point us to any information we may have missed out within the next working week when the summary needs to be finalized and published

Thanks in advance

Best regards

Name

**d) A note to communicate the findings and issuing recommendations**

Dear *Project Leader Name

Following our email exchanges I am enclosing the final summary of the findings of the knowledge audit performed under KAF methodology, as well as some recommendations based on the evaluation of your findings in relation to good knowledge sharing practices.

As part of this project we develop instruments and methods to maximize knowledge sharing and innovation in the field of systems engineering, and we would be very happy to advise you and your team further

We would welcome your feedback on your experience working with KAF, and your suggestions on how to improve the framework for future reference.

Yr Name

ENCL 1 Summary Findings
ENCL 2 Recommendation from Best Practices

# ANNEX 3

**I**n the following Table, where alternative KAF inventory template labels are adopted, they are mapped to Dublin Core standard

| | |
|---|---|
| Label: Title<br>Element Description: The name given to the resource. Typically, a Title will be a name by which the resource is formally known.<br><br>Label: Subject and Keywords<br>Element Description: The topic of the content of the resource. Typically, a Subject will be expressed as keywords or key phrases or classification codes that describe the topic of the resource. Recommended best practice is to select a value from a controlled vocabulary or formal classification scheme.<br><br>Label: Description<br>Element Description: An account of the content of the resource. Description may include but is not limited to: an abstract, table of contents, reference to a graphical representation of content or a free-text account of the content.<br><br>Label: Resource Type<br>Element Description: The nature or genre of the content of the resource. Type includes terms describing general categories, functions, genres, or aggregation levels for content. Recommended best practice is to select a value from a controlled vocabulary (for example, the DCMIType vocabulary ). To describe the physical or digital manifestation of the resource, use the FORMAT element.<br><br>Label: Source<br>Element Description: A Reference to a resource from which the present resource is derived. The present resource may be derived from the Source resource in whole or part. Recommended best practice is to reference the resource by means of a string or number conforming to a formal identification system.<br><br>Label: Related<br>Element Description: A reference to a related resource. Recommended best practice is to reference the resource by means of a string or number conforming to a formal identification system.<br><br>Label: Coverage<br>Element Description: The extent or scope of the content of the resource. Coverage will typically include spatial location (a place name or geographic co-ordinates), temporal period (a period label, date, or date range) or jurisdiction (such as a named administrative entity).<br><br>Label: Creator<br>Element Description: An entity primarily responsible for making the content of the resource. Examples of a Creator include a person, an organization, or a service. Typically the name of the Creator should be used to indicate the entity.<br><br>Label: Publisher<br>Element Description: The entity responsible for making the resource available. Examples of a Publisher include a person, an organization, or a service. Typically, the name of a Publisher should be used to indicate the entity.<br><br>Label: Contributor<br>Element Description: An entity responsible for making contributions to the content of the resource. Examples of a Contributor include a person, an organization or a service. Typically, the name of a Contributor should be used to indicate the entity.<br><br>Label: Rights Management<br>Element Description: Information about rights held in and over the resource. Typically a Rights element will contain a rights management statement for the resource, or reference a service providing such information. Rights information often encompasses Intellectual Property Rights (IPR), Copyright, and various Property Rights. If the rights element is absent, no assumptions can be made about the status of these and other rights with respect to the resource.<br><br>Label: Date<br>Element Description: A date associated with an event in the life cycle of the resource. Typically, Date will be associated with the creation or availability of the resource. Recommended best practice for encoding the date value is defined in a profile of ISO 8601 [Date and Time Formats, W3C Note, http://www.w3.org/TR/NOTE- datetime] and follows the YYYY-MM-DD format.<br><br>Label: Format<br>Element Description: The physical or digital manifestation of the resource. Typically, Format may include the media-type or | |
| dimensions of the resource. Examples of dimensions include size and duration. Format may be used to determine the software, hardware or other equipment needed to display or operate the resource.<br><br>Label: Resource Identifier<br>Element Description: An unambiguous reference to the resource within a given context. Recommended best practice is to identify the resource by means of a string or number conforming to a formal identification system. Examples of formal identification systems include the Uniform Resource Identifier (URI) (including the Uniform Resource Locator (URL), the Digital Object Identifier (DOI) and the International Standard Book Number (ISBN).<br><br>Label: Language<br>Element Description: A language of the intellectual content of the resource. Recommended best practice for the values of the Language element is defined by RFC 3066 [RFC 3066,http://www.ietf.org/rfc/rfc3066.txt] which, in conjunction with ISO 639 [ISO 639, http://www.oasis- open.org/cover/iso639a.html]), defines two- and three-letter primary language tags with optional subtags. Examples include "en" or "eng" for English, "akk" for Akkadian, and "en-GB" for English used in the United Kingdom.<br><br>Label: Audience<br>Element Description: A class of entity for whom the resource is intended or useful. A class of entity may be determined by the creator or the publisher or by a third party.<br><br>Label: Provenance<br>Element Description: A statement of any changes in ownership and custody of the resource since its creation that are significant for its authenticity, integrity and interpretation. The statement may include a description of any changes successive custodians made to the resource.<br><br>Label: Rights Holder<br>Element Description: A person or organization owning or managing rights over the resource. Recommended best practice is to use the URI or name of the Rights Holder to indicate the entity.<br><br>Label: Instructional Method<br>Element Description: A process, used to engender knowledge, attitudes and skills, that the resource is designed to support. Instructional Method will typically include ways of presenting instructional materials or conducting instructional activities, patterns of learner-to-learner and learner-to-instructor interactions, and mechanisms by which group and individual levels of learning are measured. Instructional methods include all aspects of the instruction and learning processes from planning and implementation through evaluation and feedback.<br><br>Label: Accrual Method<br>Element Description: The method by which items are added to a collection. Recommended best practice is to use a value from a controlled vocabulary.<br><br>Label: Accrual Periodicity<br>Element Description: The frequency with which items are added to a collection. Recommended best practice is to use a value from a controlled vocabulary.<br><br>Label: Accrual Policy<br>Element Description: The policy governing the addition of items to a collection. Recommended best practice is to use a value from a controlled vocabulary. | |

**ANNEX 4**

TO BE COMPLETED

Glossary of Terms

The definitions below identify specific meanings attributed to common terms within the context of the knowledge Audit Framework.

| | |
|---|---|
| Knowledge Audit Framework | A framework developed modelled on the JISC-funded DAF project to identify knowledge resourcesheld within organizations, wand to explore how they are managed. with emphasis on shareability and reuse. |
| | |
| knowledge asset | Knowledge resource that can be commercially exploited via leveraging intellectual property |
| knowledge resources | Broad terms to cover all 'public' knowledge artefacts published by an organization, |
| inventory | A detailed list of knowledge resources created by and/or used within an organization |
| registry | An online system to collect audit results |
| | |
| | |